\documentstyle[12pt]{article}
\normalbaselines
\begin{document}
\author{Branislav Jur\v{c}o \\
Department of Optics, Palack\'y
University\\V\'\i de\v{n}sk\'a 15,\\
CS-77146 Olomouc, Czech Republic\thanks{Permanent address}\\
and\\ ASI TU Clausthal, 3392 Clausthal-Zellerfeld, Germany
\thanks{Humboldt research fellow}\\}
\title{More on quantum groups from the quantization point
of view}
\date{January 18, 1993}
\maketitle
\begin{abstract}

Star products on the classical double group of a simple Lie group
and on corresponding
symplectic grupoids are given so that the
quantum double and the "quantized tangent bundle" are obtained
in the deformation description. "Complex" quantum groups
and bicovariant quantum Lie algebras are discused from
this point of view. Further we discuss the quantization of the Poisson
structure on symmetric algebra $S(g)$ leading to the
quantized enveloping algebra $U_{h}(g)$
as an example of biquantization in
the sense of Turaev. Description of $U_{h}(g)$
in terms of the generators of the bicovariant differential calculus
on $F(G_q)$ is very convenient
for this purpose. Finally we interpret in the deformation
framework some well known
properties of compact quantum groups as simple
consequences of corresponding properties of
classical compact Lie groups. An analogue of the classical Kirillov's universal
character formula is given for the unitary irreducible
representation in the compact case.
\end{abstract}
\section{Introduction}
Let $g$ be a complex simple finite-dimensional Lie algebra.
According to the Drinfeld's theorem \cite{DrQH}
(proposition 3.16) there exist a special element
${\cal F}\in (U(g)\otimes U(g))[[h]]$ such that the linear
space $U(g)[[h]]$ can be
equipped with the structure of the quasitriangular Hopf algebra,
with the standard multiplication and counit induced from $U(g)$ and
with the twisted comultiplication $\Delta_{h}$ and antipode $S_{h}$ given
by formulas
\begin{equation}
\Delta_{h}={\cal F}^{-1}\Delta {\cal F},\hskip 1cm S_{h}=u(S)u^{-1},
\end{equation}
with
\begin{equation}
u=\sum {\cal F}^{-(1)}(S{\cal F}^{-(2)}),
\hskip 1cm u^{-1}=c^{-1}\sum(S{\cal F}^{(1)}){\cal F}^{(2)}.
\end{equation}
$\Delta$ and $S$ in the above formulae are the standard
comultiplication and antipode
in $U(g)$, $c$ is a central element in $U(g)[[h]]$.
The formulas for the antipode together
with the shorthanded notation
${\cal F}=\sum {\cal F}^{(1)}\otimes {\cal F}^{(2)}$ and
${\cal F}^{-1}=
\sum {\cal F}^{-(1)}\otimes {\cal F}^{-(2)}$ are taken from \cite{Maj}.

The universal ${\cal R}$-matrix making the above Hopf
algebra a quasitriangular one is expressed with help of
symmetric $g$-invariant element $t\in g\otimes g$ (defined
via the inverse of the Killing matrix) as
\begin{equation}
{\cal R}={\cal F}^{-1}_{21}exp(ht){\cal F}.
\end{equation}

Moreover Drinfeld's theorem is claiming that this quasitriangular
Hopf algebra is isomorphic to the famous quantum group corresponding
to the Lie algebra $g$ as it was introduced by Drinfeld and
Jimbo \cite {D},\cite{J}. Let us denote by $\varphi$ this
isomorphism $\varphi : U_{h}(g) \rightarrow U(g)[[h]]$.

Comparing (1) with the explicit formula for the antipode
from \cite{RTF}
and remembering that the $\varphi$ is identity on the Cartan subalgebra
we realize that the $u$ should be proportional up to some
central element to the $\gamma =exp(-h\rho)$, where $\rho$ is the
element of the Cartan subalgebra corresponding to
the half-sum of positive roots.

In the compact case ($h\in R$) $\varphi$ can be taken to be a $\ast$-
homomorphism
\begin{equation}
\varphi (x^{\ast})= (\varphi (x))^{\ast},
\end{equation}
where the * on the left is the usual one in $U_{h}(g)$
and the * on the right is the usual one in the $U(g)[[h]]$.

Unfortunately there is no explicit formula for the {\cal F}.

In this paper we will be interested also in the dual (in the
sense of the Hopf algebras)
situation, which is nicely described in \cite{TNL}. Roughly
speaking in this situation we have on the vector space $C^{\infty}(G)[[h]]$
with the standard comultiplication and counit of the function
Hopf algebra $F(G)$
on the corresponding Lie group $G$, but the deformed
multiplication $\star _{h}$ (star product) and antipode $S_{h}$.
The corresponding
formulas expressing those with help of undeformed ones $m$ and
$S$ are:
\begin{equation}
a\star _{h}b = m({\cal F}*(a \otimes b)*{\cal F}^{-1})
\hskip 1cm S_{h}a = S(u^{-1}*a*u),
\end{equation}
for any $a,b\in F(G)$. Here the $\ast$ have been used to denote
the actions of $U(g)$ on $F(G)$ via left and right
invariant differential operators.

In the compact case we have also
\begin{equation}
a^{\ast} = \overline {\gamma * a * \gamma ^{-1}}=
\gamma^{-1} * \bar a *\gamma.
\end{equation}
Arising Hopf algebra should be of course isomorphic to the
Hopf algebra
of quantized functions $F(G_{q})$ \cite{RTF} on the group $G$ under
the $\varphi^{*}$ dual to the $\varphi$. Reader can found
an explicit example of $SL(2)$ in \cite{TNL}. We will in the
following often not distinguish between isomorphic objects under
these isomorphisms. We hope it will be clear from the context
what we have in mind.

In the above situation
the Hopf algebras $F(G_{q})$ are deformation quantizations
$(\frac{1}{h}[.,.]\rightarrow i\{.,.\}$ if $h\rightarrow 0)$
of the corresponding Poisson-Lie Groups \cite {D}, with the
Poisson bracket
\begin{equation}
i\{ a,b\} = -m(r*(a\otimes b)-(a\otimes b)*r),
\end{equation}
with $r\in g \otimes g$ being the classical $r$-matrix.

In this paper we will use this Drinfeld's idea to introduce
the star product on the classical double group of a simple Lie group
and on the corresponding
symplectic grupoids \cite{STS},\cite{Lu},\cite {LW}
to obtain the
quantum double and the "quantized tangent bundle"
in the deformation framework described above.
We will relate our results with some recent
papers on "complex" quantum groups \cite {PW},
\cite{Munich},\cite{Podles},\cite{Watamura}
and bicovariant quantum Lie algebras \cite{Z1}.

Further we will discuss the quantization of the Poisson
structure on $S(g)$ leading to the $U_{h}(g)$
as an example of biquantization in
the sense of Turaev \cite {Tu1}, \cite {Tu2} which is dual
to this of \cite {Tu2}. It turns out the description of $U_{h}(g)$
in terms of the generators of the bicovariant differential calculus
on $F(G_{q})$ \cite{Jur},\cite{Munich1},\cite{Zum} is very convenient
for this purpose.

Finally we hope to show that the deformation
reinterpretation of the group algebra of the compact
quantum group of Podles and Woronowicz \cite{PW}
lead us to some kind of universal formula for character generalizing
the classical situation \cite{Kirillov}. Here it means that
a very simple change in the classical formula allowes us to express
the trace of a functional $b\rightarrow h_{c} (ab)$
$(h_c$ is the Haar measure on $F(G_q)$, $a,b\in F(G_q)$)
in the unitary irreducible
representation of $U_{h}(g)\sim U(g)[[h]]$ in terms of
an integral on the classical coadjoint orbit with the standard
Kirillov-Souriau-Kostant symplectic structure (which is
isomorphic as a symplectic manifold to the dressing orbit
with the symplectic structure induced from the Poisson-Lie structure
on the dual group $G_{r}$ \cite{GW}). It supports the recent idea of
Xu and Weinstein \cite{XW} to construct an symplectic counterpart
of the Reshetikhin-Turaev construction of link invariants \cite{R},
\cite{RT}.

Through the paper we assume $q=e^{h}$ with generic values
of $q$.
\section{Double}
Here we will be interested in the local double Lie group corresponding
to the connected complex simple Lie group $G$. As it follows from
\cite{STS} this can be described as
$D = G\times G$. The starting group $G$ and their connected
dual group $G_{r}$ \cite{STS}
are identified with the subgroups of $D$ via its Iwasawa decomposition
\cite{HL}, which leads to the identification $D = G \times G_{r}$,
where now $G$ is assumed as the diagonal subgroup
$\{(x,x),x\in G\}$ and $G_{r}= AU^{+}$, where
$A=\{(x,x^{-1}),x\in H\}$ and $U^{+}=\{(x_{+},x_{-}),
x_{+}\in N^{+}, x_{-}\in
N^{-}\}$, with $H$ being the Cartan subgroup and $N^{\pm}$
the connected nilpotent subgroups corresponding to the
positive and negative roots respectively.
Let us remember that the double $D$ as well as the groups $G$
and $G_{r}$ are Poisson-Lie groups and that the above factorization
based on the Iwasawa decomposition is a Poisson mapping.
We will not go to the further details (reader can for more
detailed exposition consult \cite{STS},\cite{LW},\cite{Lu}).
Here we will only write down the corresponding Poisson
brackets using the classical $r$-matrix.

The Poisson
structure on $G$ have been already given by (7). We will
take the classical universal $r$-matrix $\in g
\otimes g$ of the form
\begin{equation}
r=-(P_{-} - P_{+} + t),
\end{equation}
where $P_{\pm}$ are projectors on the nilpotent subalgebras corresponding
to the positive and negative roots respectively. Here the
correspondence between the elements of $g\otimes g$ and maps
$g\rightarrow g$ via the dualization of the first factor in
$g\otimes g$ with help of the Killing form have been used.

We will
not distinguish between the universal $r$-matrix and its
representative in the fundamental representation. The
Poisson bracket (7) then becomes on the matrix elements of
the fundamental representation $\tau$
\begin{equation}
i\{\tau\otimes, \tau\} = [r,\tau\otimes \tau].
\end{equation}

The Poisson bracket in $G_{r}=\{(g_{+},g_{-})\}$ then can
be given as
\cite{BB} (we take a slightly different convention)
$$i\{g_{\pm}\otimes, g_{\pm}\} =[r,g_{\pm}\otimes g_{\pm}],$$
\begin{equation}
i\{g_{-} \otimes, g_{+}\} = [r,g_{-} \otimes g_{+}].
\end{equation}
Because of the factorizability of simple Lie algebras in
the sense of \cite{RS} the above Poisson Lie group $G_{r}$
can be as manifold identified with $G$ taken with a new Poisson
bracket. If we denote now as $y=g_{+}g_{-}^{-1}$ the
corresponding element of $G$ we have
\begin{equation}
i\{y\otimes,y
\} =  -r_{21}(y\otimes y) + (y\otimes y)r_{12}
-(y\otimes1)r_{12}(1\otimes y) + (1\otimes
y)r_{21}(y\otimes 1).
\end{equation}

The double $D$ as a Poisson manifold is a direct product of
Poisson manifolds $G$ and $G_{r}$. The Poisson  structure on the $D$ if
this is described as a manifold $G\times G$ can be also easily described
for $(\tau,\hat {\tau}) \in G\times G$
$$i\{\tau\otimes, \tau \} = [r,\tau \otimes \tau],$$
\begin{equation}
i\{\hat {\tau} \otimes,\hat {\tau}\} = [r, \hat {\tau}\otimes \hat {\tau}],
\end{equation}
$$i\{\hat {\tau}\otimes,\tau \} = [r, \hat {\tau}\otimes \tau].$$
{}From these explicit formulas it is seen immediately that
the above mentioned Iwasawa decomposition
\begin{equation}
(\tau,\hat {\tau}) = (\check {\tau},\check {\tau})(g_{+},g_{-})
\end{equation}
is a Poisson mapping.

Let us now assume the function algebra $F(D) \sim
F(G)\otimes F(G)$ with the Lie-Poisson structure given above.
If ${\cal F}\in (U(g)\otimes U(g))[[h]]$ is this one
introducing on $F(G)$ the structure of the quantum group
$F(G_q)$ with
${\cal R}$ the corresponding ${\cal R}$-matrix, then
the ${\cal F}^{D} \in (U(g\otimes g)\otimes U(g\otimes g))[[h]]$
given by
\begin{equation}
{\cal F}^{D}_{1234} = {\cal F}_{13}{\cal F}_{24}{\cal R}_{23}
\end{equation}
plays the same role for $F(D)$.
The corresponding ${\cal R}^{D}$ is
\begin{equation}
{\cal R}^{D}_{1234} = ({\cal
F}^{D}_{1234})^{-1}exp(h(t_{13}-t_{24})){\cal F}^{D}_{1234}=
{\cal R}_{41}^{-1}{\cal R}_{42}^{-1}{\cal R}_{13}{\cal R}_{23}.
\end{equation}
Here we have used the fact \cite{STS} that in the situation
of the double the $t$ from the Drinfeld's construction
should be taken as $(t_{13}-t_{24})$.

The corresponding $u$ in the formula for the antipode can
be taken as
\begin{equation}
u_{12}^{D}= {\cal R}_{21}(\gamma \otimes \gamma).
\end{equation}
In the above formulas we assume of course that the multiplication,
comultiplication, antipode and counit of the classical
double are the standard one of the direct product.

For the proof of these facts it is enough to
note that the above formulae are nothing else as a direct
application of Theorem 2.9. of \cite {RS} to the $\varphi (U_{h}(g))$.

The deformed multiplication $\star _{D}$ can be written as
$$(a\otimes 1)\star _{D} (b\otimes 1) = (a \star _{h} b)\otimes 1,$$
$$(1\otimes a)\star _{D} (1\otimes b) = 1 \otimes (a\star _{h} b),$$
\begin{equation}
(a\otimes 1)\star _{D} (1\otimes b) = a\otimes b,
\end{equation}
$$(1\otimes a)\star _{d}(b\otimes 1) = \sigma ({\cal R}*(a\otimes
b)*{\cal R}^{-1}).$$
Here $\sigma$ denotes transposition.

The * is given simply by
$$ (a\otimes b)^{\ast} = b^{\ast} \otimes a^{\ast},$$
where * on the components on the right is this one
given by (6).

Comparing (17) with the Poisson structure on the classical
double given by (12) we see that
this product is really a deformation product along this
Poisson bracket. We denote the obtained quantum double as $D_{q}$.

It is also immediately seen that the mapping $p_G=\star _{h}: F(D_{q})
\rightarrow F(G_{q})$ is a Hopf algebra homomorphism as it
should be. Similarly there is a Hopf algebra homomorphism
$p_U :F(D_{q}) \rightarrow U_{h}(g)^{op}\sim U(g)[[h]]^{op}$
:$(a\otimes b) \rightarrow {\cal R}_{21}^{-1}(.,b){\cal
R}(.,a)$, where $"op"$ means the opposite multiplication.
So factorizing out the kernels of these surjective homomorphisms
we obtain the $F(G_{q})$ and $U_{h}(g)^{op}$ as
Hopf-subalgebras.

It is instructive write out the commutation relations
resulting from (17) for the
representations of the double ${\cal T}=(\tau \otimes 1)$ and
$\hat {\cal T}=(1 \otimes \tau)$, where $\tau$ is the fundamental
representation of $G$. We have of course the familiar relations
of \cite{Munich},\cite{Podles},\cite{Watamura}
$$R{\cal T}_{1}{\cal T}_{2} = {\cal T}_{2}{\cal T}_{1}R,$$
\begin{equation}
R\hat {\cal T}_{1}{\cal T}_{2}={\cal T}_{2}\hat {\cal
T}_{1} R,
\end{equation}
$$R\hat {\cal T}_{1}\hat {\cal T}_{2} = \hat {\cal
T}_{2}\hat {\cal T}_{1}R.$$
Here $R$ is the universal ${\cal R}$ -matrix for $g$ in the
fundamental representation. We have also omitted the $\star
_{D}$ as a
sign for the multiplication.

To define the quantum double with the help of generators
${\cal T}$ and $\hat {\cal T}$ as it was done in above
mentioned papers we should suppose on the matrices
${\cal T}$ and $\hat {\cal T}$ the quantum determinant conditions
in the $A_{n}$ case and quantum orthogonality conditions in
the remaining cases. They as well as the antipode can be
obtained also from the deformation formulas. The comultiplication
and counit on ${\cal T}$ and $\hat {\cal T}$ are clearly the standard one
\cite{RTF}.

Let us now briefly discuss the quantum Iwasawa decomposition.
Let $T$ denote the matrix of generators
of $F(G_{q})$ (the fundamental representation). So we have the
famous relations of \cite{RTF}
\begin{equation}
RT_{1}T_{2} = T_{2}T_{1}R.
\end{equation}
Let $\Lambda ^{\pm}$ are the same as $L^{\pm}$ of \cite{RTF}
but now taken with the opposite multiplication
\begin{equation}
R\Lambda ^{\pm}_{1}\Lambda ^{\pm}_{2} = \Lambda
^{\pm}_{2}\Lambda ^{\pm}_{1}R,
\hskip 1cm
R\Lambda ^{-}_{1}\Lambda ^{+}_{2}= \Lambda ^{+}_{2}\Lambda ^{-}_{1}R.
\end{equation}
With help of generators $T$, $\Lambda ^{\pm}$ (entries of
$T$ are supposed to commute with those of $\Lambda ^{\pm}$)
there can be introduced according the
\cite{PW},\cite{RS},\cite{JS} another description of the
quantum double.

A slight generalization of Proposition 4.5 of \cite{JS}
shows that this Hopf algebra can be assumed as a completion of
the Hopf algebra generated by ${\cal T}$ and $\hat {\cal T}$.
The explicit formulas
\begin{equation}
{\cal T} = T\Lambda ^{+}, \hskip 1cm \hat {\cal T} =
T\Lambda ^{-}
\end{equation}
describe an injective Hopf algebra morphism $(p_G\otimes
p_U)\Delta$ of the algebra generated by ${\cal T},\hat
{\cal T}$ into the algebra generated by $T$ and $\Lambda^{\pm}$. So they give
the quantum generalization of the Iwasawa decomposition
(13). There is also the opposite Iwasawa decomposition
\begin{equation}
{\cal T} = \Lambda ^{+}T, \hskip 1cm \hat {\cal T}=\Lambda ^{-}T.
\end{equation}
Moreover the Poisson-Lie structure on
$G_{r}$ given by equations (10) can be understood as
the Poisson limit from the Hopf algebra structure of the
Hopf algebra $U_{h}(g)^{op}$ generated by $\Lambda^{\pm}$.
For the later convenience we will discuss this now for the
Poisson bracket (10) taken with the minus sign. Taking
on $G_{r}$ the coordinates $y$ we can identify $G$ and
$G_{r}$ as manifolds (locally). Now using results of
\cite {Majid Braided} we can introduce on the function algebra
$F(G)$ a new structure of an associative noncommutative algebra
expressed in the terms of the $\star _h$-product (5) as
\begin{equation}
a\star _{r}b = a_{2}\star _h b_{3} {\cal R}(a_{1},b_{2})
\tilde {\cal R}(a_{3},b_{1}).
\end{equation}
Here we used the notation $\Delta a = a_{1}\otimes a_{2}$
etc. $\tilde {\cal R}$ denotes an element of $(U(g)\otimes U(g))[[h]]$
inverse to the ${\cal R}$ but now in the algebra taken with the
opposite multiplication in the first factor. It follows from
\cite {Majid Braided} that such algebra structure is isomorphic
to the usual on of $U_{h}(g)$. Taking the classical limit
we arrive to the Poisson structure (10) with the minus
sign.
\section{Quantization of $\pi_{\pm}$}
Let us remember \cite{STS} that having a general
Lie-Poisson group $G$
with the Poisson structure given by (7) we can introduce
on the manifold $G$ two new Poisson
structures denoted by $\pi_{\pm}$.
They are is given by
\begin{equation}
\pi_{\pm}(da,db) =i \{a,b\}_{\pm}=\pm m(r*(a\otimes b) + (a\otimes b)*r).
\end{equation}
Of course $G$ equipped with these Poisson brackets is no more
a Lie-Poisson group. Nevertheless the natural left (for the +sign)
and the right (for the -sign) group actions of $G$ (G is assumed to be
equipped with the Lie-Poisson structure (7)) are
Poisson mappings.

In the case
of $G=D$ these Poisson structures are nondegenerated
\cite{STS} and the manifold $D$ with the above Poisson structures
contains all ingredients to be a symplectic grupoid over
$G$ (in the +sign case) or over $G_{r}$ (in the -sign case)
\cite{Lu}. We should warn the reader that we use different notation
as it is used in this reference.

Now making the necessary (but straightforward changes) in the
proofs of corresponding Propositions of \cite{TNL}
concerning the quantization of Poisson-Lie structure
(7) leading to the (5) we can easy state the following:

1. Formulas
\begin{equation}
a\star _{+} b = {\cal F}_{21}*(a\otimes b)*{\cal F}^{-1}
\end{equation}
and
\begin{equation}
a\star _{-} b = {\cal F} *(a\otimes b)*{\cal F}_{21}^{-1}
\end{equation}
define associative products on $F(G)$, which are quantizations of (24).

2. If we denote as $F(G,\star _h)$ and $F(G,\star _{\pm})$ the
function algebras
equipped with corresponding products, $F(G,\star _h)$ having the standard
comultiplication, then the following $F(G,\star _h)$-coactions
\begin{equation}
\delta_{+} : F(G,\star _{+}) \longrightarrow F(G,\star _h)\otimes
F(G,\star _{+}):
(\delta_{+}a)(g,h) = a(gh),
\end{equation}
\begin{equation}
\delta_{-} : F(G,\star _{-}) \longrightarrow F(G,\star _{-})\otimes
F(G,\star_ h): (\delta_{-}a)(g,h) = a(gh)
\end{equation}
are algebra morphisms.

We can now apply all above to the double $D$, with ${\cal
F^{D}}$ given by (14). The resulting algebra structure
on ${\cal T} = \tau \otimes 1$ and $\hat {\cal T} = 1 \otimes
\tau $ ($\tau$ is again the fundamental representation of $G$) gives
e.g. in the "+" case
$$R{\cal T}_{1}{\cal T}_{2} = {\cal T}_{2}{\cal T}_{1} R_{21},$$
\begin{equation}
 R\hat {\cal T}_{1}{\cal T}_{2} = {\cal T}_{2}\hat
{\cal T}_{1} R^{-1},
\end{equation}
$$ R\hat {\cal T}_{1} \hat {\cal T}_{2} = \hat {\cal
T}_{2} \hat {\cal T}_{1} R_{21}.$$
Here we have also omitted $\star _{+}$ as a sign of the multiplication.

The coaction $\delta_{+}$ is given as
\begin{equation}
\delta_{+}({\cal T})={\cal T}\otimes.{\cal T} \hskip 1cm
\delta_{+}(\hat {\cal T})=\hat {\cal T}\otimes.\hat {\cal T}
\end{equation}
with a proper understanding of the algebra structure of the
factors in the tensor products (we did not graphically distinguish
between doubles taken with different algebra structures).

The "-"case can be treated similarly.

In this same sense as in the previous section we have the
Iwasawa decomposition
\begin{equation}
{\cal T} = TL^{+}, \hskip 1cm \hat {\cal T} = TL^{-},
\end{equation}
which lead us to the following well known commutation relations
of \cite{Zum}
$$ RT_{1}T_{2} = T_{2}T_{1}R,$$
$$ R_{21}L^{\pm}_{1}L^{\pm}_{2} = L^{\pm}_{2}L^{\pm}_{1}R_{21},$$
\begin{equation}
R_{21}L^{+}_{1}L^{-}_{2}=L^{-}_{2}L^{+}_{1}R_{21},
\end{equation}
$$L^{+}_{1}T_{2} = T_{2}R_{21}L^{+}_{1}, \hskip 1cm
L^{-}_{1}T_{2} = T_{2} R_{12}^{-1}L_{1}^{-}.$$
As shown also in \cite{Zum} the relations in the last line are equivalent
to the standard pairing between $F(G_{q})$ and $U_{h}(g)$,
which is in this way implicitly contained in the quantization
of $\pi _{+}$. The action of $U_{h}(g)$ on $F(G_{q})$
resulting from these relations is easy recognized as the
left action $Xa = X*a=X(a_{2})a_{1}$ of $x\in U_{h}(g)$ on the $a\in F(G_{q})$.
Similar algebra was also introduced and investigated in
\cite{Toy model}.

We have of course also the opposite Iwasawa decomposition
\begin{equation}
{\cal T} = \tilde L^{+}\tilde T, \hskip 1cm
\hat {\cal T} = \tilde L^{-} \tilde T.
\end{equation}
The corresponding commutation relations are
$$R_{21}\tilde T_{1} \tilde T_{2} = \tilde T_{2} \tilde T_{1}R_{21},$$
$$R\tilde L_{1}^{\pm} \tilde L_{2}^{\pm} = \tilde
L_{2}^{\pm} \tilde L_{1}^{\pm}R,$$
\begin{equation}
R \tilde L_{1}^{-} \tilde L_{2}^{+} = \tilde L_{2}^{+}
\tilde L_{1}^{-}R,
\end{equation}
$$ \tilde T_{1} \tilde L_{2}^{+} = \tilde
L_{2}^{+}R_{12}^{-1}\tilde T_{1}, \hskip 1cm
\tilde T_{1}\tilde L^{-}_{2} = \tilde L_{2}^{-}R_{21}\tilde T_{1}.$$

Now computing the commutation relations
between $T$ and
$X=TL^{+}(L^{-})^{-1}T^{-1}$
we get
\begin{equation}
T_{2}X_{1} = R_{21}^{-1}X_{1}R_{12}^{-1}T_{2}.
\end{equation}
This as easy to seen is the same as the commutation relation
between $T$ and $S(L^{+})L^{-}$ if they are assumed in the algebra
which is obtained from $F(G_{q})$ and $U_{h}(g)$ as a
semidirect product with help of the right action of $U_{h}(g)$
on $F(G_{q})$ ($aX =a*X=X(a_{1})a_{2}$) \cite {Majid Braided}.
Comparing two above Iwasawa decompositions gives
$$ X={\cal T}\hat {\cal T}^{-1} = \tilde L^{+} (\tilde L^{-})^{-1}$$
and we realize that the subalgebra generated by $\tilde
L^{\pm}$ can be identified with the algebra of right-invariant maps
on $F(G_{q})$ (let us remember the known fact that the
matrices of generators $\tilde L^{\pm}$ are uniquely obtained from
the $X$ via decomposition to the triangular parts \cite{Bur}).

Let us now assume the left coaction $\delta$ of $F(G_{q})$
on $F(D,\star _{+})$ given by $\delta_{+}$ followed with the projection
$F(D,\star _h) \rightarrow F(G_{q})$ in the firts factor. As a product of two
algebra morphisms it is again an algebra morphism.
The explicit formula reads
\begin{equation}
\delta(T)=T\otimes. T, \hskip 1cm \delta(\hat {\cal
T}^{-1}{\cal T} = (L^{-})^{-1}L^{+}) = 1\otimes (L^{-})^{-1}L^{+}.
\end{equation}
As it should be $\delta$ acts trivially on left-invariant
maps.

Computing the coaction on right-invariant maps
we have
\begin{equation}
\delta ((\tilde L^{+} (\tilde L^{-})^{-1}))_{ij} =
T_{ik}T_{lj}^{-1} \otimes (\tilde L^{+} (\tilde L^{-})^{-1})_{kl},
\end{equation}
which is nothing else as the left dressing action of $F(G_{q})$
on $U_{h}(g)^{op}$ corresponding to the opposite Iwasawa
decomposition of the double given by (22) \cite{JS}.

With a slightly different conventions the above coaction $\delta$
have been investigated also in \cite{Z1}. Here we hope it
was introduced in a more general context.

We will now briefly discuss some facts generalizing the
classical situation \cite {STS},\cite{Lu}. Owing to the
above described Iwasawa decompositions of $F(D,\star _{+})$
we have the following natural projections (algebra homomorphisms)
$$p_{1}:F(D,\star _{+})\longrightarrow F(G_{q}),$$
\begin{equation}
p_{2}:F(D,\star _{+})\longrightarrow U_{h}(g)
\end{equation}
corresponding to the first Iwasawa decomposition (31)
and
$$\tilde p_{1}:F(D,\star _{+})\longrightarrow F(G_{q})^{op},$$
\begin{equation}
\tilde p_{2}:F(D,\star _{+})\longrightarrow U_{h}(g)^{op}
\end{equation}
corresponding to the opposite one (33).

Further as easily shown by direct calculation the entries
of the matrices ${\cal T}\hat {\cal T}^{-1}$ (right-invariant
elements) and
$\hat {\cal T}^{-1}{\cal T}$ (left-invariant elements)
as it should be mutually commute.
So the subalgebras of $F(D,\star _{+})$ generated by $L^{\pm}$
and $\tilde L^{\pm}$ can be viewed as a generalization of
the notion of the dual pair from the classical case. As in
the classical case \cite{W} their only common elements are
their Casimirs.

It is well known \cite{STS}, \cite{LW}, \cite{Lu}, that the
symplectic manifolds $(D,\pi_{\pm})$ play an important role
in the description of symplectic leafs of the Poisson
structures on $G$ and $G_{r}$. Their quantization presented in
this Section should play an analogous role in the representation
theory of $F(G_{q})$ and $U_{h}(g)$. E.g. the irreducible finite-
dimensional representations of $U_{h}(g)$ for the generic
value of $q$ can be obtained decomposing the
representation of $U_{h}(g)$ on $F(G_{q})$ given by the left
action $X*a, X\in U_{h}(g), a\in F(G_{q})$ which is contained
as it was shown in the
algebra structure of the $F(D,\star _{+})$.

We hope that also in the quantum case all ingredients to satisfy
the formal definition of quantum grupoid \cite{Compt.rend}
are contained in $F(D,\star _{\pm})$.

\section{Biquantization of $S(g)$}

This section is motivated by appendix of \cite{Tu2}, where
the dual situation have been described.
Here we will use without further explanation the
terminology introduced in \cite{Tu1},\cite{Tu2}. The reader
is referred to the algebraic parts of this papers. We think
that it is not necessary to reproduce here all details, because
we hope that the presented example is enough illustrative. Minor
differences from \cite{Tu1},\cite{Tu2} are insubstantional.

Let us remember that a simple Lie algebra $g$ can be
equipped in a standard way with a structure of Lie bialgebra
\cite{D}. It means that in addition to the Lie bracket $[.,.]$
we have also a Lie cobracket $\nu :g\rightarrow g\wedge
g$ (which is according to \cite {D} equivalent to the
Poisson structure on $G$) and the Lie bracket and Lie cobracket
are compatible. In our
case the Lie cobracket is given in terms of the classical
$r$-matrix
\begin{equation}
\nu (X)=[r,1\otimes X+X\otimes 1],\hskip 1cm X\in g.
\end{equation}
The symmetric algebra $S(g)$ is then endowed with
the structure of so called biPoisson bialgebra. Roughly speaking
it is equipped in addition to the commutative
multiplication and the standard Poisson bracket (given by
the extension of the Lie bracket on $g$ via the Leibniz rule)
with the

1. comultiplication $\Delta$ (coalgebra structure)
\begin{equation}
\Delta (X) = 1\otimes X + X\otimes 1,\hskip 1cm  X\in g,
\end{equation}
which is extended to the entire $S(g)$ as an algebra
homomorphism.

2. Lie cobracket (coPoisson structure)
given on $g$ by (40) and extended to the entire
$S(g)$
with help of the rule
\begin{equation}
\nu (ab) = \nu (a) \Delta (b) + \Delta (a) \nu (b).
\end{equation}

Let us note that using the classical Yang-Baxter equation
we can write the Poisson bracket in coordinates $X =
(\tau\otimes id)(r-r_{21})$ ($\tau$ -the fundamental representation,
$r$-the classical universal $r$-matrix) as
\begin{equation}
i\{X_1 , X_2 \} =  [r+r_{21}, X_2 ].
\end{equation}
It is a well known fact \cite{LW},\cite{STS} that this
Poisson structure is the linearization of the Poisson
structure on $F(G_{r})$.

\vskip 0.5cm
Further let $V$ be the associative algebra over $C[h]$
obtained in the following way:

It is the tensor algebra $T(C[h]\otimes g)$ over
$C[h]\otimes g$ divided by two-side
ideal generated by elements of the form
$$ ab-ba-h[a,b]. $$ The generators of $V$ are simply only
rescaled generators of $g$.
We have $V/hV = S(g)$ and $V/(h-1)V=U(g)$.
In this same way as above we can equip $V$ with a
coalgebra and coPoisson structures.
The $V$ as an $C[h]$-algebra is quantization of Poisson algebra
$S(g)$ in the usual sense. The
corresponding projection
$q_{h}:V\rightarrow S(g)$ is called quantization
homomorphism. $q_{h}$ is of course a surjective bialgebra
homomorphism and preserves the cobracket.
If $r$ is the classical $r$-matrix (8), then
collecting the generators of $g$ with the help of the fundamental
representation $\tau$ into the matrix $X = (\tau \otimes id)(r-r_{21})$
we can write thanks to the classical Yang-Baxter equation
the commutation relation in the form
\begin{equation}
X_{1}X_2 -X_2 X_1 = h[r+r_{21},X_2].
\end{equation}
\vskip 0.5cm

Let us now assume the $U_{h.\hbar}(g)$. It means that in the
definition of the quantized enveloping algebra we make simply
the change $h\rightarrow h.\hbar$, where $\hbar$ is a new
Planck constant. Further let us
introduce with the help of the standard $R$-matrix a new matrix $\bar R$
 \begin{equation}
 \bar R  (h\hbar)= (h\hbar)^{-1}(R(h\hbar)-I)
 \end{equation}
so that we have
\begin{equation}
\bar R =-r +O(h\hbar).
\end{equation}
We will collect the generators in the matrix denoted again as
$X$

\begin{equation}
X = (\hbar)^{-1}(L^+ S(L^-) - I).
\end{equation}
We have following commutation relations
$$(X_1 X_2 - X_2 X_1) + h\hbar(\bar R_{21}X_1 X_2 +X_1 \bar
R_{12}X_2- X_2 \bar R_{21}X_1 -X_2 X_1 \bar R_{12})$$
$$+ (h\hbar)^2 (\bar R_{21} X_1 \bar R_{12}X_2 - X_2 \bar
R_{21}X_1 \bar R_{12})$$
\begin{equation}
= -h^2 \hbar (\bar R_{21} \bar
R_{12}X_2 -X_2 \bar R_{21}\bar R_{12}) - h[\bar R_{12}+
\bar R_{21}, X_{2}].
\end{equation}
The comultiplication is given as
\begin{equation}
\Delta (X_{ij}) = X_{ij}\otimes 1 + 1\otimes X_{ij} +
X_{kl}\otimes (L^+_{ik} S(L^-)_{lj} -\delta_{ik} \delta_{jl}).
\end{equation}
Here it is assumed that the entries of $L^{\pm}$ in the
last term are expressed as functions of $X$, which is
possible due to the already mentioned triangular decomposition
of $\hbar X +I$.
The entries of the matrix $X$ are nothing else as properly
normalized generators of the bicovariant differential
calculus on $F(G_{q=exp(h\hbar)})$
\cite{Jur},\cite{Munich1},\cite{Zum}.
The $C[[h,\hbar]]$-bialgebra $A$ generated by $X$ (47) with
relations (48-49) is
essentially the $U_{h}(g)$, which is a completion of
$A/(\hbar -1)A$.

In the limit $\hbar \rightarrow 0$ we get the above described
co-Poisson algebra $V$ (more precisely their completion via
the inclusion $C[h]\hookrightarrow C[[h]]$) so that $A$ is
a co-qauntization of $V$
in the sense that for $a\in A$ holds
\begin{equation}
(p_{\hbar}\otimes p_{\hbar})\hbar ^{-1}(\Delta(a)-\sigma
\Delta(a)) = \nu (p_{\hbar}(a)).
\end{equation}
The corresponding projection $p_{\hbar}$, which is again a
surjective bialgebra homomorphism, is a called
co-quantization homomorphism.
\vskip 0.5cm

The last what we need in our discussion of the
biquantization of $S(g)$ is the following Poisson
bialgebra $F$. Let $y=g_+g_-^{-1}$ be the coordinates in $G_{r}$
introduced in the Section 1. Let us introduce new coordinates
collected in the matrix again denoted as $X$ via the relation
$X = \hbar ^{-1} (y-I)$. Poisson bracket in this new
coordinates reads
\begin{equation}
i\{X_1 ,X_2 \}= \hbar (r_{21}X_1 X_2 + X_1 r X_2 -X_2 r_{21}
X_1 -X_2 X_1 r) + [r+r_{21}, X_2].
\end{equation}
The comultiplication is given by
\begin{equation}
\Delta (X_{ij}) = X_{ij}\otimes 1 + 1\otimes X_{ij} +
X_{kl}\otimes ((g_+)_{ik}(g_-)^{-1}_{lj} -\delta_{ik} \delta_{jl}).
\end{equation}
The $g_{\pm}$ are assumed as functions of $X$ in the last term.
The Poisson bialgebra $F$ over $C[\hbar]$ with
coordinates $X$ is nothing else as $F(G_{r})$. Namely
the later is a completion of $F/(\hbar-1)F$.

We also immediately see that $A$ is a quantization of $F$
(in the limit $h\rightarrow 0$ we get the completion
of $F$ via the inclusion $C[\hbar] \hookrightarrow C[[\hbar]])$.
$F$ self is a co-quantization of $S(g)$. The
corresponding surjective bialgebra homomorphisms $p_{h}$
and $q_{\hbar}$ are thus quantization and co-quantization
homomorphisms respectively. Moreover the map $q_{\hbar}$ is
an Poisson algebra homomorphism.
\vskip 0.5cm
The collection consisting of the bialgebra $A$, co-Poisson
bialgebra $V$, Poisson bialgebra $F$ , homomorphims $p_{h}$,
$p_{\hbar}$, $q_{h}$, $q_{\hbar}$ and the surjective
bialgebra homomorphism
$$p: A\longrightarrow S(g),$$
\begin{equation}
p=q_h \circ p_{\hbar}= q_{\hbar} \circ p_{h},
\end{equation}
which is simultaneously the quantization of the  Poisson
bracket and co-quantization of the co-Poisson bracket in
$S(g)$ realize the notion of the reduced
biquantization of the bi-Poisson
bialgebra $S(g)$.
\vskip 0.5cm
Now we will briefly discuss some consequences of the facts
collected above in the context of the
representation theory. We will be interested only in
the algebraic and Poisson structures appeared (we will
forget all co-algebra and co-Poisson algebra structures).
Let us remember that $S(g)\sim Pol(g^*)$ and
let us assume a particular
integral coadjoint orbit $O$ in $g^*$ of the maximal
dimension (we assume in
the following the connected and simply connected, simple
compact groups ). According to the classical
Borel-Weyl-Bott theorem there is one to one correspondence
between such orbits and irreducible unitary
representations. The quantization homomorphism $q_{h}$
"restricted" to the particular representation $T_O$ (which
extends to be an irreducible unitary representation of $V$) and the
corresponding orbit $O$ (we hope that it is clear what we mean)
is most conveniently described with the help of coherent states
connected with $T_O$ \cite{Per}. The
introduction of the Planck constant $h$ in the commutation
relations as it was done in the case of introducing the
algebraic structure of $V$ is a common trick used for the
discussion of the classical limit \cite{Yaffe}.
The corresponding modification of coherent states described
there is this just we need. We refer the reader to this
article for more detailed discussion. In our situation we
know that the range of the exponential mapping is up to a
set of a zero measure the entire $G$. If $\Lambda_i$ are the
generators of $g$, than the corresponding coherent states
are defined as
\begin{equation}
e^{\frac{1}{h} \vec {\lambda}.\vec {\Lambda}}\vert >,
\end{equation}
where the coefficients $\lambda_i$ are taken such that the
exponent belongs to the definition domain of the
exponential mapping and the $\vert >$ is the highest weight
vector of $T_O$. Now as easy to see the covariant symbols
of rescaled generators $h\Lambda_i$ have the proper limit
as $h\rightarrow 0$ and the resulting symplectic
structure is the usual Kirillov-Souriau-Kostant one as we
need.

The algebra homomorphism $p_{\hbar}$ is described with the
help of the algebra isomorphism $\varphi$.
Formula
$$\hbar ^{-1}((\tau \otimes id){\cal R}(h\hbar)_{21} {\cal R}(h\hbar)
 - I) = X$$
gives an expression of the generators $X$ in the terms of
the generators of $g$, which are nothing else as
$h^{-1}\times$generators of $V$. This way this formula
gives also the expression of $X$ in terms of generators of $V$.
The projection $p_{\hbar}$ is then given by taking the
limit $\hbar \rightarrow 0$ in the last expression.

So the discussed  representation extends to the unitary
irreducible representation of $A$ which as already
discused is essentially the quantized enveloping algebra
of $g$. The representative operators of $X_{ij}$ are now
quantizations of their covariant symbols \cite{Berezin}
in the coherent states representation. The limit
$h\rightarrow 0$ from the symbols gives of course
a Poisson map $F\rightarrow O$ and describes such a way
a symplectic leaf of $F$. As well known \cite
{STS},\cite{LW} the last should be a dressing orbit of $G$ in $G_{r}$.
It is also known \cite{GW} that the Poisson manifolds $g^*$
and $G_{r}$ are isomorphic and that this isomorphism sends
the coadjoint orbits into the dressing orbits. So
we have finally arrived to a particular realization
of this isomorphism. It is clear that there is a similar
relation between this isomorphism and the algebra
homomorphism $q_{\hbar}$ as it was between $\varphi$ and
$p_{\hbar}$.

\section{Compact quantum groups and the trace formula}

Now we will discuss some consequences of introducing the
*-structure in the way described in Section 1 (6).

Following the lines of proofs of Proposition 3.16 of \cite{DrQH}
and Proposition 4.3 of \cite{D1} it is easy to seen that in
the compact case ${\cal F}{\cal F}^{\ast}$ is $g$-invariant. The * in
$U(g)[[h]]\otimes U(g)[[h]]$ is the usual component-wise one.
{}From the property of the ${\cal R}$-matrix (Proposition 4.2
of \cite{D1})
$${\cal R}^{\ast} = {\cal R}_{21}$$
we conclude the symmetry of ${\cal F}{\cal F}^{\ast}$. Twisting
\cite{DrQH} with help of symmetric $g$-invariant element
$({\cal F}{\cal F}^{\ast})^{-1/2}$ leads to the new one
$\tilde {\cal F}$ which is unitary. We will assume in the
following the ${\cal F}$ to be unitary. As a simple
consequence we have
$$u^*=S(cu^{-1}).$$
\vskip 0.5cm
Further if $t^{\alpha}$ denotes the unitary irreducible
representation of $G$, then computing the
$(t^{\alpha}_{ij})^{\ast}$ and $S_h (t^{\alpha}_{ji})$ according to
the formulas (6) and (1). It means $t^{\alpha}$
is also a unitary irreducible representation of the $G_{q}$
(corepresentation
of the $F(G_{q})$ ) in the sense of \cite{W1} (the
comultiplication remains unchanged). So the Peter-Weyl
theorem generalizes immediately from the classical to the
quantum case. The last is of course also well known
\cite{W1}.  The above
$\star _{h}$-product is in this case the same as introduced
in \cite{DV} and the Weyl transformation described there
is identical with isomorphism $\varphi^*$. The Haar measure
$h_c$ on the $G_{q}$ of \cite{W1} reduce under this isomorphism
to the usual Haar measure $\eta= \int dg$ on $G$ as
already noted in \cite{DV}.

Using the definition of the $\star_h$-product and the
property of the Haar measure $\eta$
\begin{equation}\eta ((x*a)b)= \eta (a (S(x)*b)),
\hskip 1cm \eta
((a*x)b) = \eta (a(b*S(x))),
\hskip 1cm a,b\in
F(G), x\in U(g),
\end{equation}
we have
\begin{equation}
\eta (a\star _h b) = \eta((S(cu^{-1})*a*u)b)= \eta
(a(((cu^{-1})*b*S(u))).
\end{equation}
Applying these formulas and the definition of *
(6) we can compute for the unitary irreducible
representations $t^\alpha$ and $t^{\beta}$ of $G$ using the well known
orthogonality relations for the compact groups \cite{Kirillov}
\begin{equation}
\eta (t^{\alpha} _{ij}\star_h (t^{\beta})^{\ast}_{kl}) = \frac{1}{dim(\alpha)}
\delta _{\alpha \beta} \delta _{ik} uS(cu^{-1})(t^a_{lj}).
\end{equation}
Comparing this with the Theorem 5.7 of \cite {W1} we see
that
\begin{equation}
dim(\alpha)\gamma ^2 = M_{\alpha} uS(cu^{-1}),
\end{equation}
where we have made an identification between $\gamma ^2$ and
$f_1$ of \cite{W1} and where $M_{\alpha}$ denotes the trace of
$\gamma ^2$ in the representation $t^{\alpha}$.

Similarly computing $\eta ((t^{\alpha})^{\ast}_{ij} \star_h t^{\beta}_{mn})$
we have
\begin{equation}
dim(\alpha)\gamma ^{-2} = M_{\alpha} cS(u)u^{-1}.
\end{equation}
Further using the two possible expressions for the $c$
following from the definitions of $u$ and $u^{-1}$ (2)
and the fact that ${\cal F}$ is now unitary we get
immediately from the above equations (57) and (58)
\begin{equation}
cc^{\ast} = \bigl(\frac {dim(\alpha)}{M_{\alpha}}\bigr)^2,
\end{equation}
valid for any unitary irreducible representation $t^{\alpha}$ of $G$.
Let us denote the positive square root of the casimir
$cc^{\ast}$ as $N$. We have $N=S(N)$ and we can rewrite the
formula (56) as
\begin{equation}
\eta(a\star_h b) = \eta (((N\gamma) *a *\gamma )b)= \eta
(a((N\gamma ^ {-1})*b*\gamma ^{-1})).
\end{equation}

Let us now remember that according to \cite{PW} we can to any
$\hat a \in F(G_q)$ relate an functional $\xi_{\hat a}$
which gives on $\hat {b}\in F(G_q)$ the value
\begin{equation}
\xi _{\hat a} (\hat b) = h_c (\hat {a} \hat b).
\end{equation}
In our deformation description we have owing to (61)
\begin{equation}
\xi _a (b) = \eta(((N\gamma) *a*\gamma )b) =\eta (a
((N\gamma ^{-1})* b*\gamma ^{-1}))
\end{equation}
for $a,b\in F(G)$.
On the other hand we have also the classical functional
$\xi ^{cl}_a$ given simply by
\begin{equation}
\xi ^{cl}_a =\eta (ab).
\end{equation}
Comparing (63) and (64) we obtain
\begin{equation}
\xi _a =N\gamma ^{-1}\xi ^{cl}_a \gamma ^{-1}=\xi ^{cl}_{\tilde
a},
\end{equation}
with
$$\tilde a =(N\gamma) *a*\gamma . $$

Let us now assume the unitary irreducible representation $t_O$
of $G$ (simply connected, connected, compact) corresponding to the
integral coadjoint orbit of the maximal dimension (it is as already
noted above a unitary irreducible representation of $G_q$).
We can now apply the classical universal trace formula \cite{Kirillov}
to the functional $\xi ^{cl}_{\tilde a}$
to compute the
trace of the operator $\xi_a$ in this representation. We get
\begin{equation}
Tr_O (\xi_a)=\frac{dim(O)}{M_O} Tr_O(\gamma^{-2} \xi
^{cl}_a)= Tr_O (\xi_{\tilde a})=
\end{equation}
$$=\frac{dim(O)}{M_O} \int _O \left(\int
_U a(\gamma exp(X)
\gamma )Q^{-1}(exp(X)) e^{2\pi i <F,X>} dX \right) d\beta_O(F).
$$
In the last formula $F \in g^*$, $<.,.>$ is the dualization
between $g$ and $g^*$, $U$ is the inverse image in $g$ of an
open region (of the complement of the zero measure) in $G$ covered by
the canonical system of
coordinates, $\beta _O$ means the canonical measure on $O$
defined by the Kirillov-Souriau-Kostant symplectic form,
$dX$ - Lebesgue measure on
$g^*$ and $Q$ is a universal function \cite{Kirillov}.
Here of course the ordering of terms in the argument of $a$
is insubstantional. We have taken the most symmetric one.

So the trace of the functional $h_{c}(\hat a.)$, $\hat a\in
F(G_{q})$ in the unitary
irreducible representation of $G_q$ is expressed in
the following way:

1.  in the Peter-Weyl expansion are matrix elements
of unitary irreducible representations of $G_q$ replaced by
those of the corresponding unitary irreducible
representations of $G$ so we get the corresponding element
$a=(\varphi ^*)^{-1}(\hat a) \in F(G)$ and

2.  applying the trace formula to the so obtained $a\in
F(G)$ to express the trace with help of integration over the
corresponding coadjoint orbit (dressing orbit) $O$.

Let us note the interesting fact that the trace of of $\xi
_a$ is proportional to the Markov trace of $\xi ^{cl}_a$.

We finish with the simple formulas  for the left and right
invariant measures $h_{dL}$ and $h_{dR}$ of \cite{PW} on
$\xi _a$.
\begin{equation}
h_ {dR}(\xi _a)=\sum _O dim(O)Tr_O(\xi ^{cl}_a) = a(e),
\end{equation}
which agrees with the formula (2.24) of \cite{PW} (here we have
instead the left invariant measure the right one, because
the $\rho _a $ of \cite{PW} differs by antipode from $\xi _a$).
For the left invariant Haar measure we get
\begin{equation}
h_ {dL}(\xi _a)=\sum _O dim(O) Tr_O (\xi
^{cl}_{\gamma ^4 *a})=a(\gamma ^4).
\end{equation}

So using the deformation formalism described above we can
view many properties of the compact quantum groups
in particular some of them described in
\cite{W1},\cite{PW} as consequences of the
well known properties of the compact Lie groups.
\section{Acknowledgement}
The author is much indebted to S. Zakrzewski, M.Schlieker
and W.Weich for numerous discussions. The idea that the
algebra defined by relations (32) is the quantization of
the Poisson structure $\pi_+$ (24) was suggested to the author by
S. Zakrzewski. The author is also indebted to Prof. H. D.
Doebner for his kind hospitality in Clausthal. The work was
supported by the Alexander von Humboldt Foundation.

\end{document}